FIGURE 1

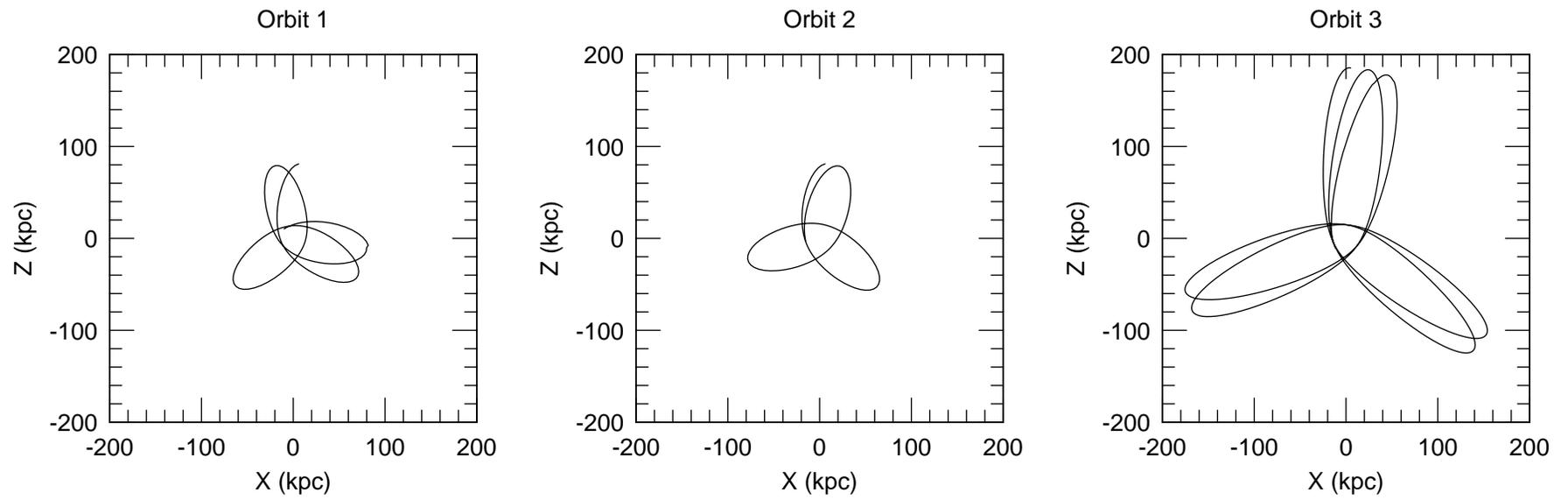

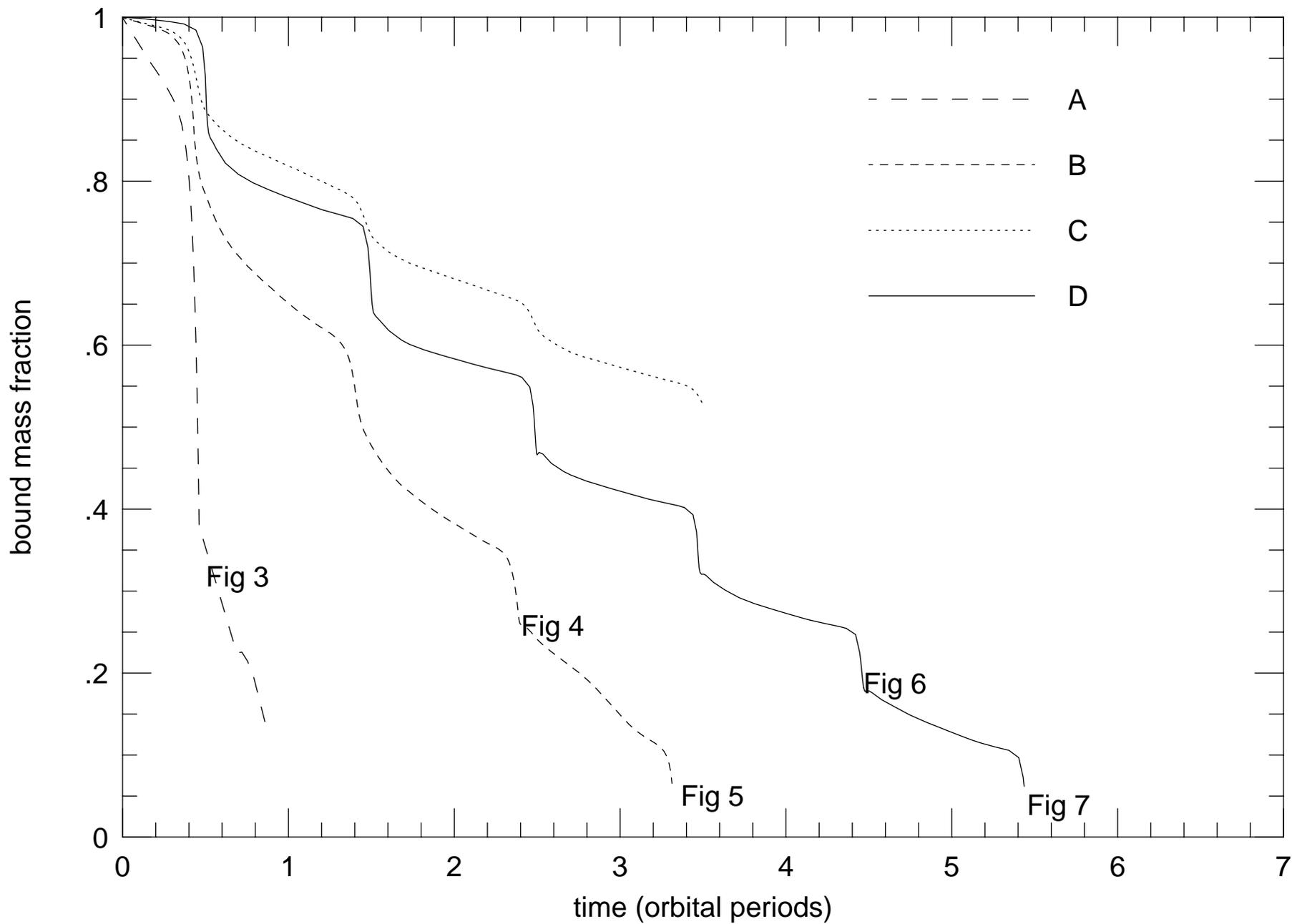
FIGURE 2

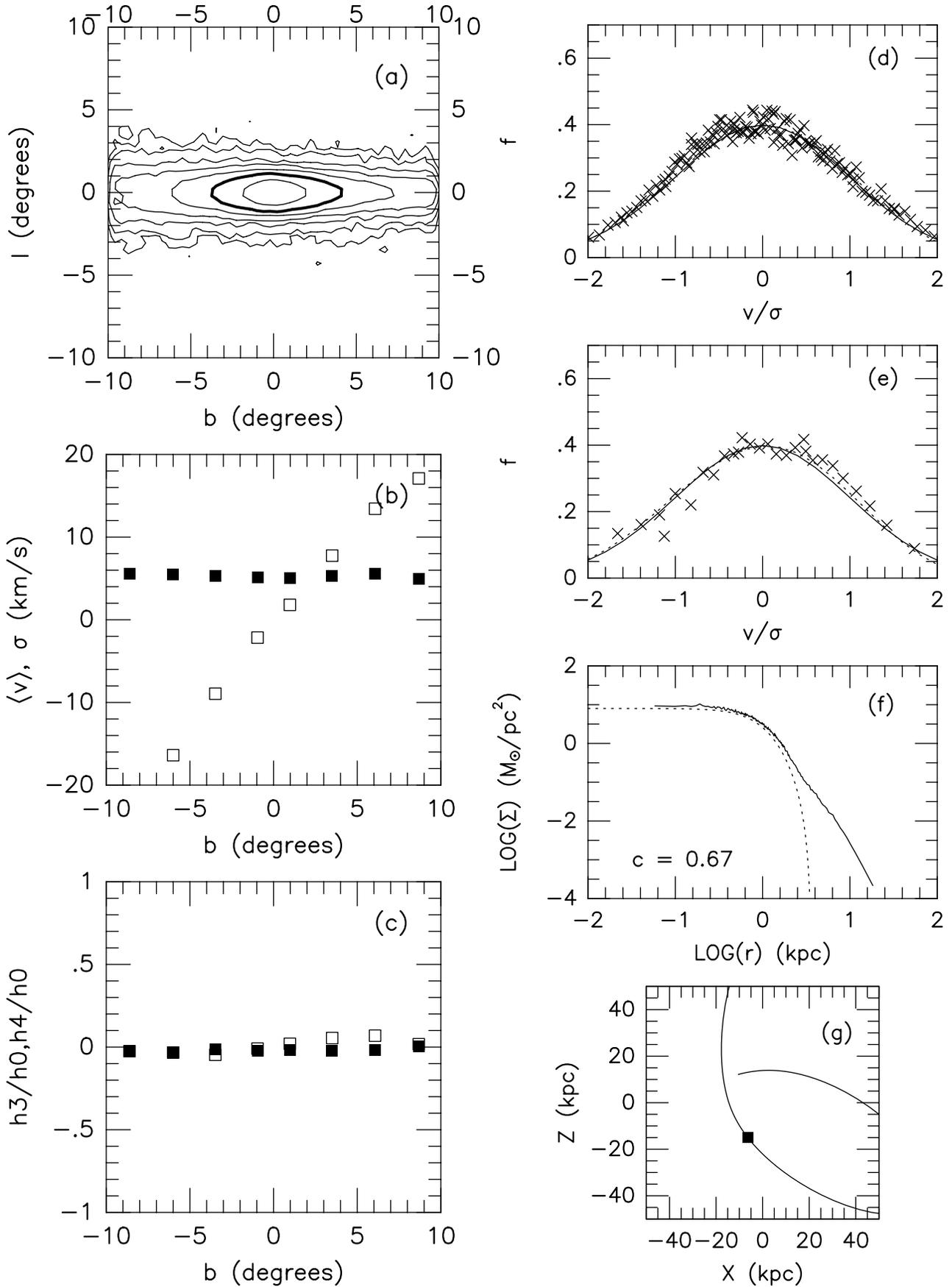

Figure 3

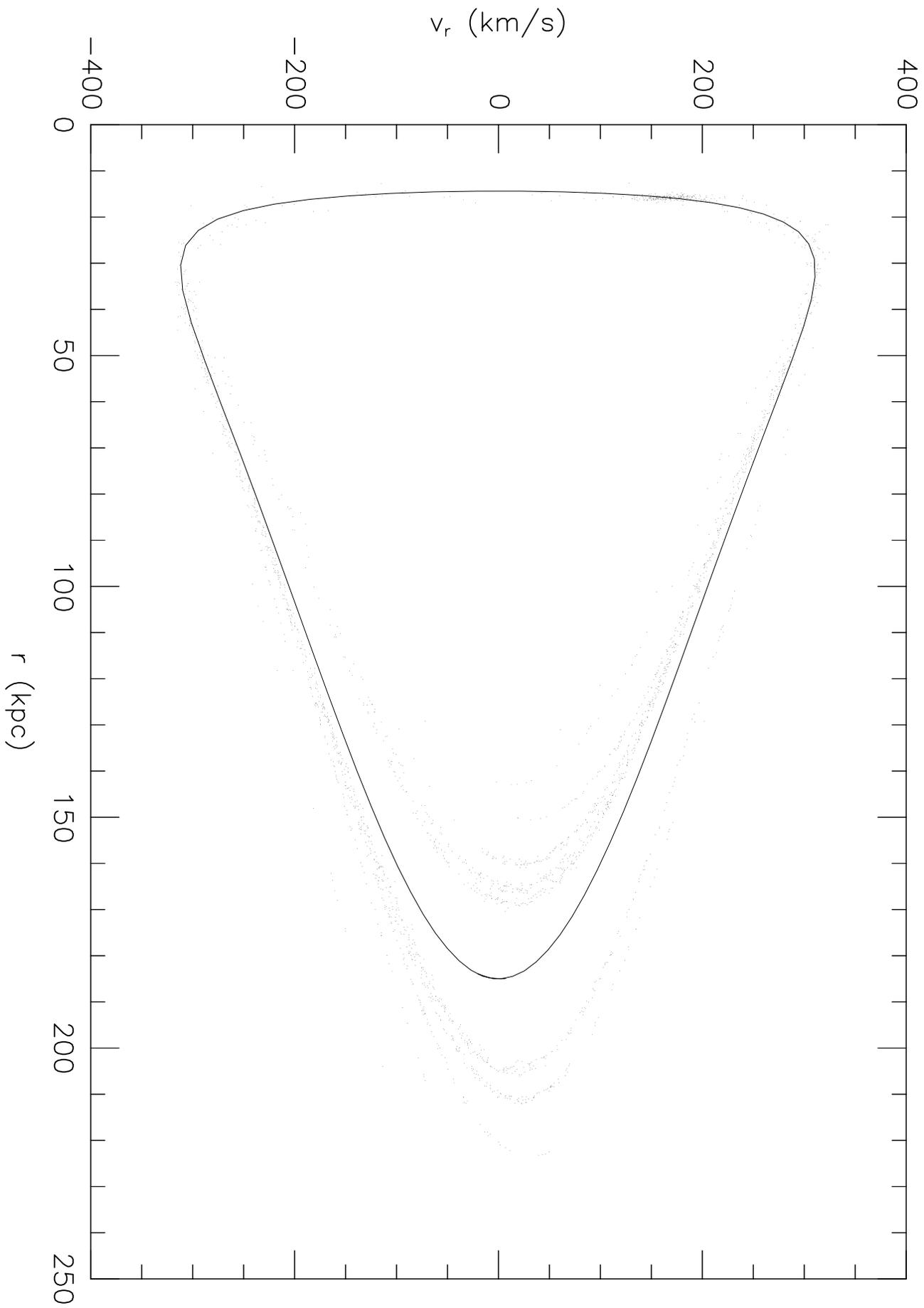

FIGURE 8

FIGURE 9

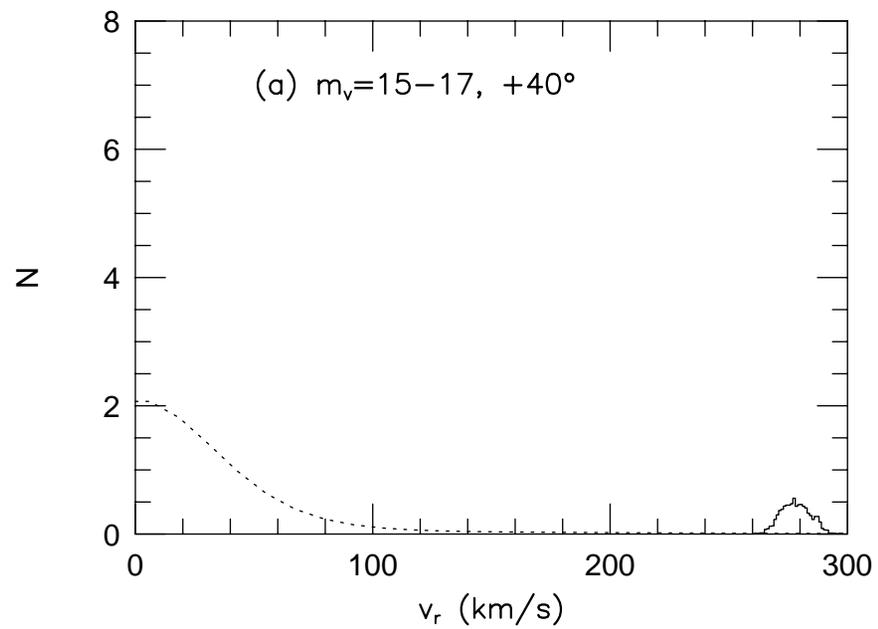
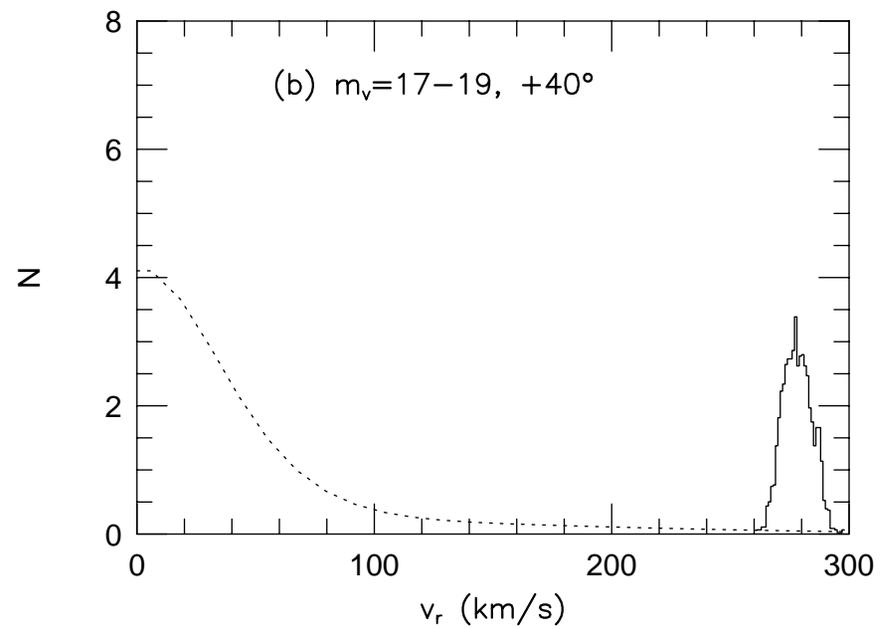
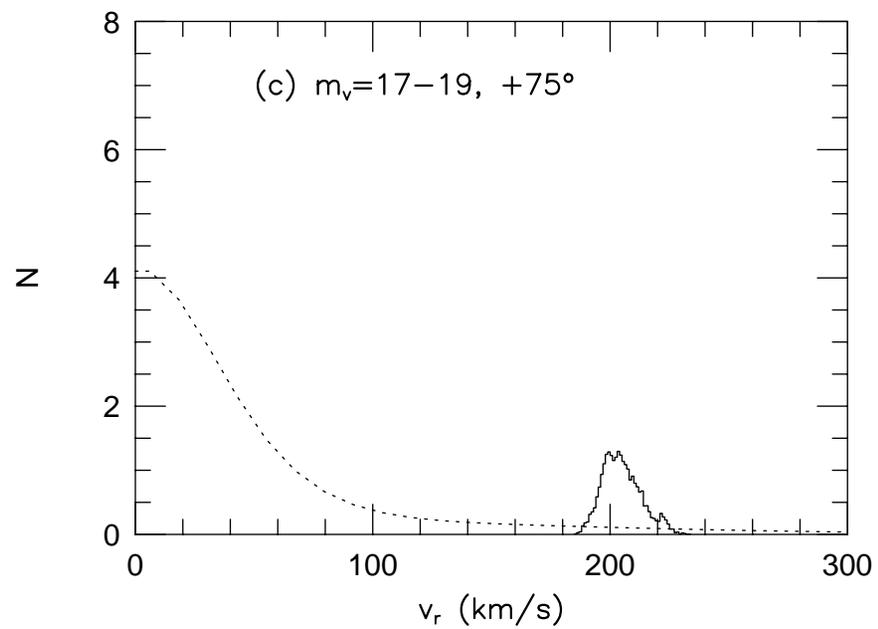
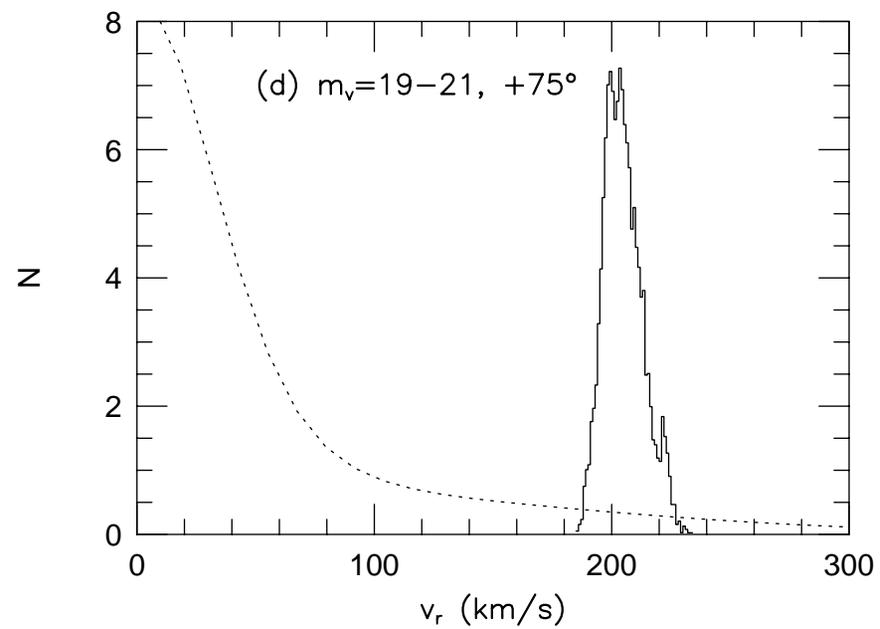

FIGURE 10

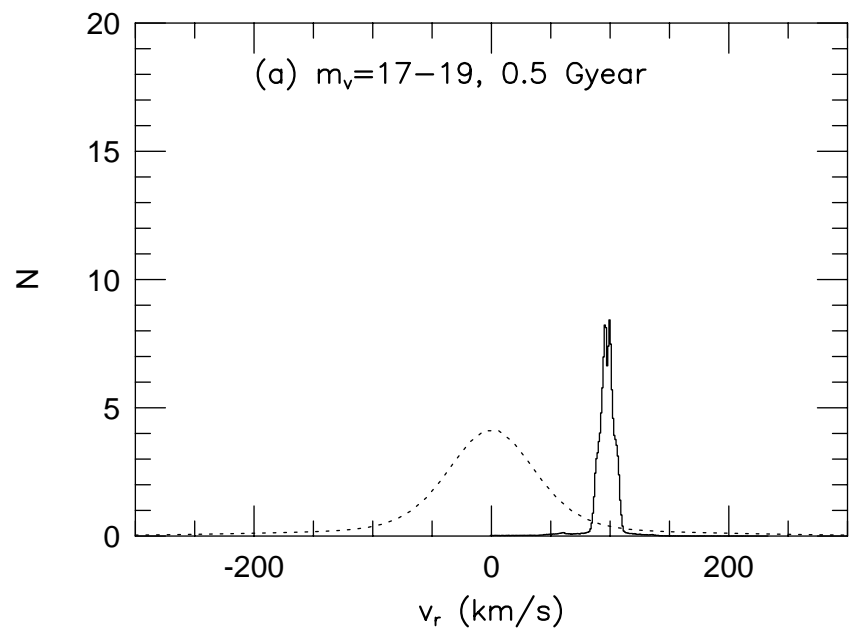
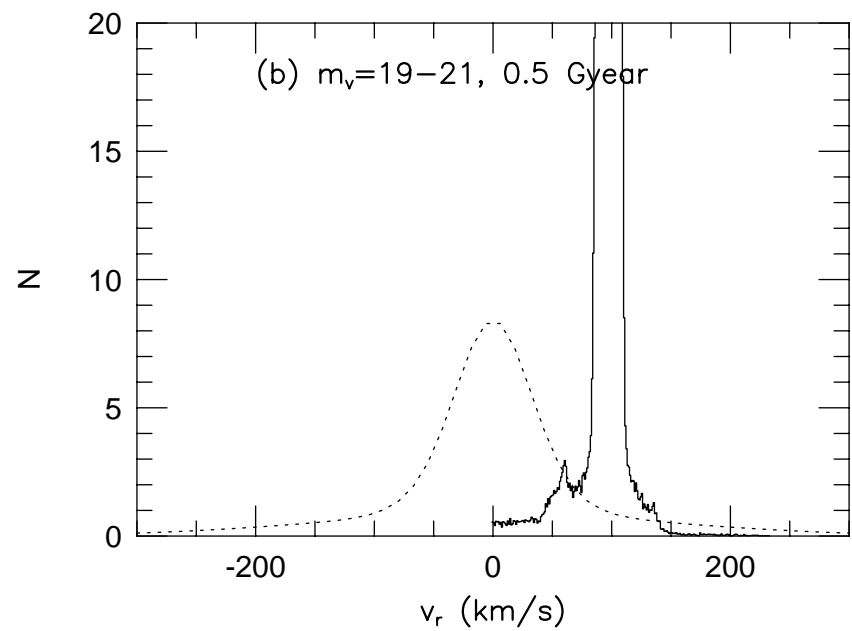
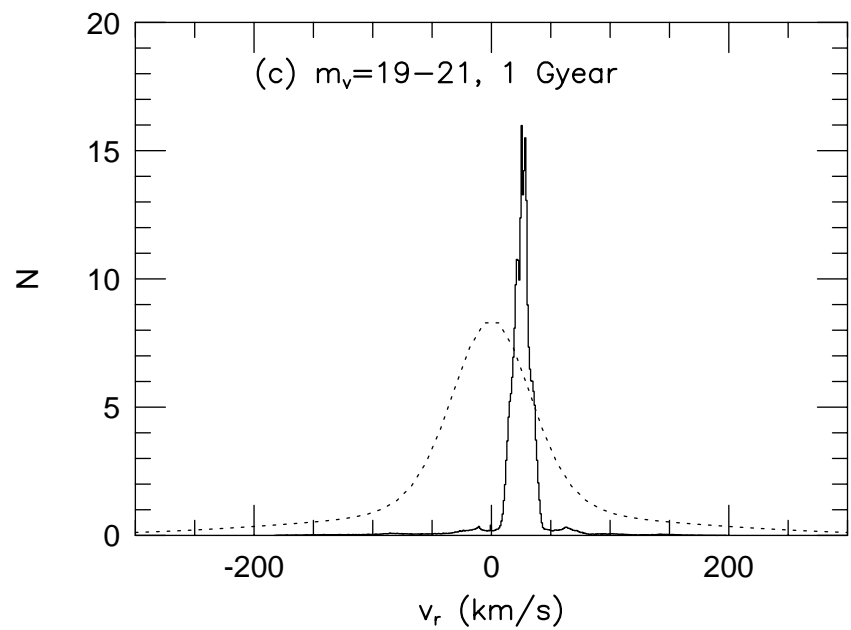
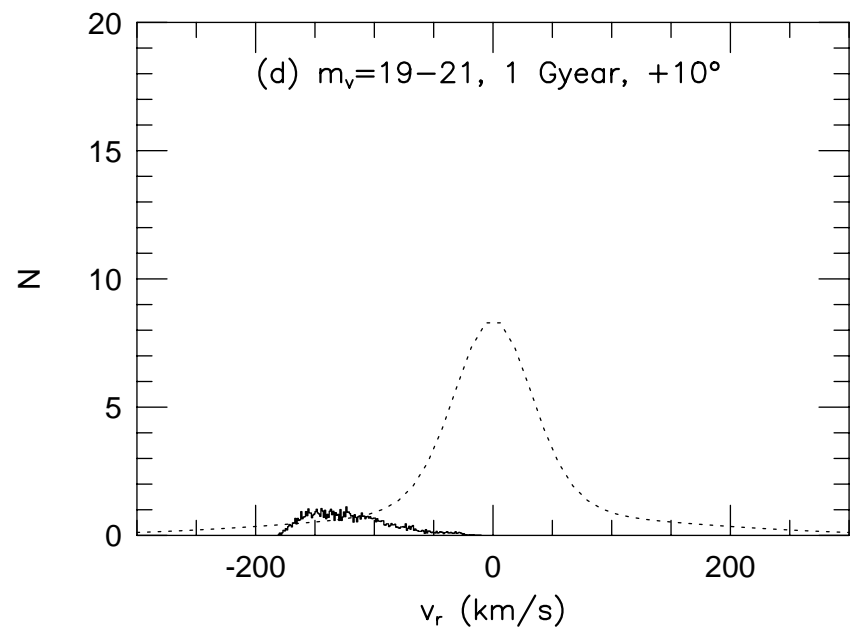

# The Disruption of the Sagittarius Dwarf Galaxy.

Kathryn V. Johnston[1], David N. Spergel[2], and Lars Hernquist[1,3].

## ABSTRACT

Numerical simulations of dwarf spheroidal galaxies undergoing several close encounters with the Milky Way are described. By comparing our models to observed properties of the recently discovered dwarf galaxy in Sagittarius (Sgr), we discuss implications of our results for the formation and evolution of the Milky Way system. We find that existing observations are not sufficient to allow us to place precise limits on either the orbit or the initial state of the dwarf. Debris from the ongoing tidal stripping of the Sagittarius galaxy are expected to form moving groups in the halo of the Galaxy and the discovery of such stars would strongly constrain the history and dynamical state of the dwarf. Furthermore, if Sgr is presently being disrupted, we predict that its remains will be detectable as a moving group in the halo for more than 1 Gyr. Thus, if similar accretion events have occurred in the recent history of the Galaxy, their aftereffects may still be observable.

*Subject headings:* galaxies - interactions, galaxies - evolution, galaxies - formation, galaxies - Milky Way

---

[1] Board of Studies in Astronomy and Astrophysics, U. C. Santa Cruz, Santa Cruz, CA 95064
[2] Princeton University Observatory, Princeton University, Princeton, NJ 08540
[3] Sloan Foundation Fellow, Presidential Faculty Fellow

## 1. Introduction

In April 1994, Ibatha, Gilmore & Irwin (1994; hereafter IGI) announced the discovery of a moving group in the direction of Sagittarius, centered on the Galactic coordinates l= 7.5°, b= −15°, and which is 24 ± 2 kpc from the Sun (∼ 16 kpc from the center of the Galaxy). Surface density contours of this object are elongated perpendicular to the disk of the Milky Way and are characterized by an axis ratio of ∼ 3 : 1. The lowest and highest contour levels, which cover the region $l = 5°$ to $10°$, $b = -20°$ to $-10°$, span a factor of ∼ 10 in surface density. The stars in the group have a heliocentric radial velocity of 153 ± 2 km/s, corresponding to 176 km/s in Galactocentric coordinates, a velocity gradient of < 1 km/s per degree along the major axis, and a velocity dispersion of 10 km/s.

Mateo *et al.* (1994) subsequently obtained deep photometry of a field in this area and estimate that members of the group have typical ages of ∼ 10 Gyrs and metallicities of [Fe/H]≈ −1.0 ± 0.3. These results, together with the lack of HI in this region (IGI), support IGI's conjecture that the moving group is a dwarf spheroidal galaxy (dSph), which would make it the ninth thus far discovered around the Milky Way. IGI compare Sgr directly to the Fornax dSph since the number of horizontal branch (HB) stars they detect (∼ 17000) suggest that the two dwarfs are similar in size, leading to a magnitude estimate of $M_v \sim -13$ (or $L_{tot} \sim 10^7 L_\odot$) for Sgr.

Sgr's current location offers a rare opportunity to study the effects of a strong tidal interaction in detail. The tidal radii of clusters of stars of masses $10^8 M_\odot$ and $10^7 M_\odot$ at 16 kpc from the center of the Galaxy are 0.77 kpc and 0.36 kpc, respectively, assuming an enclosed mass of $3 \times 10^{11} M_\odot$, whereas the group extends over ∼ 2 kpc, suggesting that the dwarf is currently being disrupted. The extreme axis ratios inferred for Sgr are a reflection not only of the compressive shock of the encounter on the dwarf but also of the dispersal of stripped material along its orbit. Thus, the direction of Sgr's elongation traces its projected orbit.

We model the dynamics of the Milky Way and Sgr dwarf by considering encounters that are consistent with the observed properties of the group. We examine whether or not existing observations are sufficient to determine the dynamical state of the group; namely, if it is bound or unbound and whether or not it has suffered previous close encounters. Since only one component of the dwarf's velocity is known, we do not know its precise orbit and cannot hope to reproduce all observations exactly, but rather must be content to investigate the general characteristics of such encounters.

The mere existence of Sgr has implications for models of galaxy formation since the timescale for the encounter is ∼ 16 kpc/176 km/s < $10^8$ years. Clearly, accretions of dwarf galaxies by the Milky Way must be common or we must instead conclude that we are observing the Galaxy during a special phase of its evolution. The former possibility supports Searle and Zinn's (1978) proposal that the Galaxy was slowly built up from primordial fragments and implies that the known dSph's represent only a small fraction of a much larger original population of satellites (for reviews see Larson [1990] and Majewski [1993a]). To investigate this possibility, we follow the evolution of one simulation beyond complete tidal disruption and discuss observable consequences of such an event.

## 2. Method

### 2.1. Models

A three-component model is used for the Galaxy, in which the disk is represented by a Miyamoto-Nagai potential (1975), the spheroid by a Hernquist potential (1990a) and the halo by a logarithmic potential:

$$\Phi_{disk} = -\frac{GM_{disk}}{\sqrt{R^2 + (a + \sqrt{z^2 + b^2})^2}},$$

$$\Phi_{spher} = -\frac{GM_{spher}}{r + c},$$

$$\Phi_{halo} = v_{halo}^2 \ln(r^2 + d^2).$$

Here, $M_{disk} = 1.0 \times 10^{11}$, $M_{spher} = 3.4 \times 10^{10}$, $v_{halo} = 212$, $a = 6.5$, $b = 0.26$, $c = 0.7$, and $d = 12.0$, where masses are in $M_\odot$, velocities are in km/s and lengths are in kpc. This choice of parameters provides a nearly flat rotation curve between 1 and 30 kpc and a disk scale height of 0.2 kpc. The radial dependence of the z epicyclic frequency ($\kappa_z$) in the disk between radii at 3 and 20 kpc is similar to that of an exponential disk with a 4 kpc scale length.

Initially, the satellite is represented by a Plummer model

$$\Phi = -\frac{GM}{\sqrt{r^2 + r_0^2}},$$

where $r_0 = 0.6$ kpc and the mass is either $M = 10^7 M_\odot$ or $M = 10^8 M_\odot$.

For $r_0 = 0.6$ kpc, a mass of $10^7 M_\odot$ yields a central density comparable to the luminous component of the Fornax dSph and a central velocity dispersion of 3.4 km/s, while the higher mass includes the full dynamical mass of Fornax (Mateo et al. 1993) and gives a central velocity dispersion of 10.7 km/s. Table 1 summarizes each of the simulations and provides analytic estimates of tidal radii (King 1962) and the impulsive energy input for each model on its first encounter (see Binney & Tremaine 1987). Note that Model D employed 50 times as many particles as the others to permit a more detailed analysis of phase space.

Three different orbits were used in the simulations. Table 2 summarizes the properties of these orbits and Figure 1 traces their paths. Each orbit is constrained to have a Galactocentric radial velocity of 176 km/s at a distance of 16 kpc from the center of the Galaxy and is chosen to lie in a plane perpendicular to the disk, in view of Sgr's direction of elongation. Orbits 1 and 3 illustrate the effect of varying eccentricity, assuming that Sgr has always been bound to the Milky Way, and are run in our full Galactic potential. An orbit much less eccentric than Orbit 1 would lead to the complete destruction of our dwarfs in unacceptably short ($< 1$Gyr) timescales, while an orbit much more eccentric than Orbit 3 would have an apocenter beyond the most distant observed dSphs (Leo I and Leo II). Orbit 2 is virtually identical to Orbit 1, but includes the effect of only the spheroid and halo potentials so that the relative influence of the disk can be determined.

### 2.2. Evolution Code

All computations were performed on the Connection Machine 5 (CM-5) at the National Center for Supercomputing Applications (NCSA). The Milky Way is represented by a rigid potential, while the dwarf is described by a collection of self-gravitating particles whose mutual interactions are calculated using a self-consistent field code (Hernquist & Ostriker 1992). Our code has been parallelized to run on the CM-5 (Hernquist, Sigurdsson & Bryan 1994) to maximize particle number and consequently minimize numerical relaxation (Hernquist & Barnes 1990, Weinberg 1993, Johnston & Hernquist 1994). Since the mass of the Sgr dwarf is much smaller than the mass of the Milky Way, dynamical friction and energy exchange between the parent and satellite galaxies are assumed negligible and have been ignored.

### 2.3. Analysis

To interpret our results, we convert the contours of the observed isopleth map of Sgr (see IGI) into mass surface densities. The lowest contour seen is at 0.125 image/(arcmin)$^2$, which corresponds to 0.0026 image/pc$^2$ at a distance of 24 kpc, where an "image" is an HB star. This number density, $n$, can be related to a mass surface density using

$$\Sigma = \left(\frac{n}{0.0026}\right)\left(\frac{1}{B}\right)\left(\frac{A}{528L_\odot}\right)\left(\frac{C}{2.25}\right) 3 M_\odot/\text{pc}^2,$$

where $A$ is the total luminosity of the system per red giant branch (RGB) star with magnitudes less then the magnitude of the HB, $B$ is the ratio of the number of HB stars to the number of RGB stars with magnitudes less than the magnitude of the HB, and $C$ is the mass-to-light ratio for the system. The magnitude of the HB can be approximated by

$$M_{HB} = 0.66 + 0.19[Fe/H]$$

(Bolte 1994), and $A$ can be calculated from an appropriate luminosity function. Using Bergbusch and VandenBerg's (1992) isochrones for an 8.0Gyr, $[Fe/H] = -1.03$ cluster, and an initial mass function exponent $x = 1.5$ gives $A = 528$. $B$ can be estimated from observed luminosity functions and ranges from $B = 1$ to $B = 2$ for various ages and metallicities (e.g. Mighell 1990; Buonanno, Corsi & Fusi Pesci 1990). $C$ is 2.25 for the chosen luminosity function. Hence we estimate that $\Sigma > 1 M_\odot/\text{pc}^2$ for the lowest contour in IGI's isopleth map.

### 3. Results

In what follows we first consider the effects of varying model parameters and then compare results to observations of Sgr. Figure 2, described in §3.1, illustrates the evolution of each model; Figures 3 – 7, described in §3.2 – §3.4, summarize the state of the models at times consistent with Sgr's current position and velocity relative to the Sun; and Figures 8 – 10, described in §3.5 – §3.7, explore the existence and persistence of moving groups in the halo.

Observations of Sgr constrain the average Galactocentric radius (16 kpc) Galactocentric radial velocity (176 km/s), and position relative to the Galactic disk (15°) of the most tightly bound particles in the simulation. To *exactly* match all these parameters on each of many passages along an orbit with a given pericenter and apocenter would require a corresponding number of simulations, with each one chosen to match the constraints on one particular passage, because the orbit is a rosette and hence the angular position of pericenter precesses. However, the halo and the disk contribute nearly equally to the disruption (see §3.1) and the dwarf receives roughly the same energy impulse on each encounter (to within a factor of two) irrespective of the exact orientation of the disk relative to the orbit or location of disk passage. Indeed, Figure 6, which illustrates the fifth passage of Model D, shows no anomalous features when compared to the other figures, even though its most recent disk passage occurred at more than 30 kpc and the impulse from the disk was significantly smaller than in the other passages. In addition, the current position, radial velocity and elongation of Sgr itself suggest that it is not far from its own pericenter and disk passage and is likely to have suffered from a similar impulse to that experienced by the models in our simulations. Hence, as a practical compromise, we drop the last observational constraint (orientation relative to the plane of the Galaxy) and, in Figures 3 – 7, consider the state of the dwarf only when it is at a radius of $\sim 16$ kpc with a radial velocity of $\sim 176$ km/s.

### 3.1. Mass Loss and Disruption

Figure 2 shows the fractional mass still bound by self-gravity as a function of time during each of the simulations. The curves clearly show that mass loss occurs primarily and almost instantaneously at pericenter owing to the impulsive energy input from the encounter. The orientation of the disk with respect to the orbit and the exact distance of the dwarf from the Galactic center as it passes through the disk, which both vary on each passage, do not affect the impulse significantly since the mass lost at pericenter is always of the same order.

By comparing Models B and D, which differ only in their orbits, we see that the mass lost at pericenter depends weakly on orbit parameters. Model B survives fewer encounters than Model D, however, because it loses mass at a greater rate throughout its (less eccentric) orbit.

Models B and C, which differ only in that the Galactic disk is excluded in Model C, show substantial differences in the rate of mass loss. This result emphasizes that rather than being merely a disk-shocking phenomenon, tidal disruption involves both disk and halo shocking, the latter being much like bulge-shocking of globular clusters (e.g. Aguilar, Hut & Ostriker 1988). In fact, the internal potential energy of bound particles in Model C after its fourth encounter is 1/3 its initial value, representing an energy gain of 2/3 its initial binding energy. Since Model B, run in the full potential, was completely disrupted over the same number of encounters, we infer that the last 1/3 of its binding energy must have been gained from the disk. We conclude that at least half of the energy gained from the encounters in the full potential comes from halo shocking alone. In addition, from the distorted shape of its surface density contours it is impossible to infer if Sgr has yet passed through the disk since the disk is not solely responsible for the satellite's destruction.

### 3.2. Surface Densities

In Figures 3 – 7 we present results from the viewpoint of an observer at rest with respect to the center of the Galaxy and 24 kpc from the dwarf on the opposite side of the bulge.

Panel (a) of each figure shows mass surface density contours, with the bold contour at the level $1 M_\odot/\text{pc}^2$. The isopleths of the observed star counts are much less smooth than the contours in our figures, probably due to uncertainties in the counts themselves and because of dust obscuration towards the bulge of the Galaxy. However, each simulated map compares well in extent, ellipticity, density contrast, and absolute density to the observations.

Panel (f) shows the surface density profile calculated on concentric ellipses, with an ellipticity chosen to fit the contour at 5° along the major axis (solid line). The dotted line is a King model of the given concentration, plot-ted for comparison. The central surface density and concentration parameter of the King model fitted to the profiles both decrease with increasing number of pericentric passages until a critical point is reached, after which full disruption occurs. On the last passage prior to disruption, all our models (irrespective of orbit and initial conditions) are well fitted by nearly identical King models. This interesting coincidence merits further investigation, but is not something we discuss further here. A fuller exploration of parameter space (varying both the mass and the scale length of the initial model) would be necessary to address this issue since the criterion for tidal disruption involves the density contrast of the satellite and parent galaxies. (Previous numerical investigations of tidal interactions and density criteria for disruption include Miller [1986], McGlynn [1990], and Oh, Lin and Aarseth [1994].)

### 3.3. Kinematics

For the kinematic analyses all velocities are projected along the line of sight noted in the previous section. Particles are ordered by their position along the major axis (positive angles are in the direction of motion of the dwarf) and then binned into eight $(1.25°)^2$ boxes before calculating the velocity moments. No bin contains fewer than 100 particles.

Panel (b) shows the average velocity (open symbols) and velocity dispersion (filled symbols) of particles in bins along the major axis of the contour map. It is interesting to note that tidal stripping does not enhance the line-of-sight velocity dispersion, even out to projected distances of $10°$.

The models do have a substantial positive velocity gradient ($> 2$ km/s per degree) along their major axes, in contrast to the observations of $< 1$ km/s per degree. Two effects contribute to line of sight velocity gradients observed along the projected major axis of a satellite in the tidal field of the Milky Way. The first is due to the combination of the tangential velocity, $v_t$, of the satellite with its angular extent on the sky. An angular separation of $\Delta\theta$ between two lines of sight along the major axis leads to a velocity difference of $v_t \sin \Delta\theta$ (assuming that the major axis is coincident with the tangential velocity vector). The second is due to stripping of material from the satellite. Stars forming the leading streamer are pulled towards the Galaxy to lower energy orbits and those forming the trailing streamer are pulled away to higher energy orbits (Oh, Lin, & Aarseth 1994). Clearly the sign and magnitude of the gradient will depend on the position of the observer relative to the Galactocentric radius vector of the satellite and the strength of the tidal interaction. In our case (for a line of sight coincident with the radius vector), the first effect leads to a positive velocity gradient in the direction of motion of the satellite, while the second leads to a negative gradient. In Figures 3 – 7 the dwarf has a large angular extent on the sky and the first effect dominates, with Model D (Figures 6 and 7) showing the most significant gradient because it is on a higher energy orbit and has a larger $v_t$. That such a gradient is not observed in Sgr could be because the two effects are of the same magnitude.

Panel (c) shows normalized values of the Gauss-Hermite moments calculated for particles in each bin. These moments quantify the deviation of the velocity distribution from a Gaussian (Gerhard 1993; van der Marel & Franx 1993). We have chosen van der Marel

& Franx's formalism using the derivation outlined in Heyl, Hernquist & Spergel (1994). Since the higher order moments are only a few percent of the zeroth order term, the distribution is nearly Gaussian along the major axis, as confirmed in panels (d) and (e) which show the velocity distribution of particles in the bins at $-9°$ and $-1°$, respectively.

### 3.4. Comparison with Observations

There are no striking differences between the results shown in Figures 3 – 7, and each fits most, but not all, of the observations well. The points of ambiguity are the smoothness of the density contours and the large velocity gradients seen along the projected major axes in our simulations. The absence of such a velocity gradient in the observations is very surprising in the light of these simulations and may provide an important clue to Sgr's history. If the lack of gradient is confirmed in further observations then it will place strong constraints on the orbit of Sgr by requiring that the two effects contributing to any velocity difference (as described in §3.3) exactly cancel.

In summary, our results show that Sgr is neither necessarily on its first nor on its last pericentric passage. There are no outstanding features in the velocity distribution to distinguish between a passage in which a bound core of particles still remains, as in Figures 4 and 6, from one in which all particles are unbound, as in Figures 5 and 7. Severely distorted contours are apparent in all these figures. Thus it is difficult to place constraints on the initial mass of the dwarf or on its orbit with current observational data.

### 3.5. Phase Space Structure

For the remainder of §3 we discuss the position-velocity phase space evolution of Model D in detail. This model employed 50 times as many particles as the others to improve resolution and accuracy. In addition, this was the only model with a disruption timescale of order the Hubble time.

Figure 8 shows the radial velocity as a function of radius for a sample of 2500 particles immediately following the final passage of the satellite (at the time illustrated in Figure 7). The solid line traces the dwarf's orbit. At this time the model has undergone 6 pericentric passages over the course of 13 Gyrs and it has been fully disrupted. Each encounter has produced its own distinct tidal streamer. Nevertheless, the particles occupy a relatively small region of phase space. We hope to exploit this high density to locate tidal debris from cannibalized dwarfs in the halo of our Galaxy.

### 3.6. Streamers as Moving Groups

The Sagittarius dwarf was first recognized as a moving group. Is it possible to detect the streamers from the encounter away from the Galactic center? In Figure 9, we examine the line-of-sight velocity distributions of stars in two different square-degree regions. The viewpoint is again that of an observer at rest with respect to the center of the Galaxy and 24 kpc from the dwarf on the opposite side of the bulge. Panels (a) and (b) are for stars $40°$ away from the dwarf along its orbit, while panels (c) and (d) show similar information at $75°$. Apparent magnitudes are estimated using the adopted luminosity function (solid line). Also plotted in each of the panels is an estimate of the observed numbers of stars as a function of line-of-sight veloc-ity in these directions (dotted line). These distributions were generated using the star counts of thin disk, thick disk, and halo stars in the direction of the North Galactic Pole (Reid & Majewski 1993). The velocity distribution of each component is modeled as a Maxwellian with z-velocity dispersions of 30, 45, and 150 km/s for the three components, respectively. A comparison of the areas under the two curves in each panel shows that the surface density of the stars in the streamers is only $\sim 10\%$ of the background. However, the tidal streamers stand out as moving groups in all magnitude bins, though the exact distribution depends on the details of the encounter. Observing such a group directly associated with the Sagittarius dwarf would strongly constrain its orbit, and hence its history. The tidal streamers are expected to trace the great circle defined by Sgr's elongation.

For Model D's orbit, however, it seems unlikely that the streamers could also be observed as moving groups in proper motion studies. The square-degree surveyed at 40° contains particles at $\sim 37$ kpc with tangential velocities of $\sim 370$ km/s, which corresponds to a proper motion of $2.2 \times 10^{-3}$ arcsecs/year, well below the current observational limits for such surveys.

### 3.7. Moving Groups following Disruption

In Figure 10 we consider the persistence of moving groups in the halo, following Model D beyond its complete destruction on its sixth pericentric passage. The model maintains a "core" of unbound stars at densities higher than the tidal streamers for more than 1 Gyr after this passage. This "core" could not be confused with a dwarf spheroidal galaxy, however, since its surface density quickly falls below $0.1 M_\odot/\text{pc}^2$, an order of magnitude smaller than that of the Galaxy's satellites (Mateo et al. 1993). The persistence of this "core" does not contradict estimates of the timescale for the dwarf to disperse following its disruption ($\sim r_0/\sigma \sim (500 \text{ pc})/(10 \text{ km/s}) \sim 10^8$ years). At 1 Gyr after disruption the dwarf's surface density has already fallen by a factor of 10 because of the free streaming of the higher velocity stars away from the region. The remaining "core" consists of the lowest velocity stars whose dispersion is much less than 10 km/s and, hence, whose dispersion timescale is much longer than $10^8$ years.

Panels (a) and (b) of Figure 10 show the velocity distribution of stars in the "core" in two different magnitude bins, 0.5 Gyrs after pericenter. Panel (c) is a similar plot, 1 Gyr after pericenter. All three suggest that such a group could be observed. Panel (d) is the velocity distribution of stars 10° further along the orbit than those in panel (c), and here the moving group is not so obvious. Note, however, that these stars are more than 160 kpc away from the center of the Galaxy. Presumably it would be less challenging to detect a disrupting dwarf on a more tightly bound orbit.

### 3.8. Carbon Stars and Globular Clusters

The Fornax dSph contains ~65 carbon stars (Richer & Westerlund 1983) and 5 globular clusters (Hodge 1988). The disruption of such a system might leave some signatures of the event as substructure in the halo carbon star and globular cluster populations. Indeed, IGI report the existence of four globular clusters in the vicinity of Sgr itself.

In the previous section, the square-degree examined in Figure 10(a) contains ~5% of the initial mass of the model. For a Fornax-like population, such a region would contain approximately 3 high-velocity carbon stars, giving a surface density significantly higher than the average (0.02 deg$^{-2}$) estimated to a depth of $M_v = 18$ for faint, high-latitude carbon stars by Green, Margon, Anderson & Cook (1994).

Clearly, the strategy of identifying debris by its over-density in a small region of phase-space would not work for globular clusters from disrupted dwarfs. However, the satellite galaxies of the Milky Way tend to lie in the great circles defined by the Magellanic Stream (the Magellanic Plane) and the Fornax-Leo-Sculptor stream, as noted first by Lynden-Bell (1976, 1982). Globular clusters have subsequently been associated with both these planes (Lin & Richer 1992; Majewski 1994). These great circles present obvious targets to search for moving groups and their discovery would be further proof that the alignment of halo objects is not mere coincidence, but rather the signature of tidal interactions and accretions of satellites by the Milky Way.

## 4. Discussion

Five main conclusions follow from our work:

1. For pericenters of $\gtrsim$ 13 kpc the contribution of the halo to the disruption of a dwarf spheroidal of mass less than $10^8 M_\odot$ is at least that of the disk.

2. The dwarf galaxy in Sagittarius is neither necessarily on its first nor on its last Galactic passage, despite tidal radius arguments.

3. From existing observations it is not possible to distinguish amongst a wide range of orbital parameters and initial dwarf models. It is also not possible to determine the direction of motion of Sgr from its distorted state and, so, we cannot say if we are currently observing it immediately prior to or following disk passage.

4. Debris from the dwarf's most recent passage should be observable as moving groups in the halo.

5. It may be possible to infer whether or not similar events have occurred within the past $10^9$ years from observations of moving groups in the halo.

Conclusion (1) has implications for models of Galaxy formation. The existence of the thin disk is often used as evidence against substantial recent accretion of satellites by the Galaxy (e.g. Tóth & Ostriker 1992; Quinn, Hernquist & Fullagar 1993, and references therein). In our simulations, if half of the initial binding energy of a $10^8 M_\odot$ satellite is converted entirely into random energy in a $10^{11} M_\odot$ disk, this would increase the z-velocity dispersion ($\sigma_z$) in the disk by only ~ 0.25 km/s. Moreover, the important contribution of the halo to tidal disruption implies that dwarfs could easily be destroyed on orbits with larger pericenters by halo shocking alone, with negligible effect on the disk.

These findings are consistent with earlier studies which stress the fragility of disks (Quinn and Goodman 1986; Quinn, Hernquist & Fullagar 1993; Walker, Mihos & Hernquist 1994). In the simulations of Walker *et al.* (1994), for example, the satellite orbit decays quickly both radially and into the plane of the disk through dynamical friction with the disk and the halo. A substantial fraction of both the orbital and internal energy of the satellite is lost to the disk and $\sigma_z$ can increase significantly through a single accretion event. In our calculations, the ratio of satellite mass to disk mass is an order of magnitude smaller than previous studies. Since frictional deceleration is proportional to the mass of the satellite ($M_{sat}$), the timescale for orbital decay ($\propto 1/M_{sat}$) is much longer (see Binney & Tremaine 1987). More of the orbital energy is carried away by stripped particles and less is deposited in the disk. Hence, our estimate of an increase of 0.25 km/s in $\sigma_z$ suggests that the Galaxy could absorb $10^9 - 10^{10} M_\odot$ in small accretion events and still maintain a thin disk. This supports Searle and Zinn's view of the formation of the Milky Way from primordial fragments. (For further discussion, see e.g. Hernquist [1990b], Hernquist & Quinn [1993], and Majewski [1993b]) However, fully self-consistent simulations are needed to precisely determine the relative importance of the disk and halo to the orbital decay and disruption of small satellites.

Conclusions (4) and (5) emphasize the value of observational kinematic studies. A number of moving groups have already been reported in the literature (see Preston, Beers & Shectman 1994; Majewski, Munn & Hawley 1994, and references therein). The discovery of moving groups in the halo associated with Sgr would constrain its orbit, its origin, and its history. Observations of such groups in general (or limits on their occurrence) would tell us something about the frequency of accretion events in the recent history of our Galaxy. A smooth phase-space distribution argues in favor of a relatively uneventful past, while a non-uniform one points to the more common occurrence of dwarf galaxy accretion, much as we are witnessing today in Sagittarius.

Since we do not know the history of the Galaxy, we cannot rely on surveys covering only a few square arc-minutes (e.g. Bahcall, Flynn, Gould & Kirhakos 1994) to tell us much beyond the local properties of the region observed. If the halo was indeed formed from many accretion events we would expect non-uniformity on square-degree scales because of this history and it would only be through a survey covering many square-degrees on the sky that we could obtain global estimates of its content and structure

We thank Mike Bolte for many helpful comments and Doug Lin, Mike Irwin, Neil Tyson, and Burt Jones for useful discussion. This work was supported in part by the National Center for Supercomputing Applications, the Alfred P. Sloan Foundation, NASA Theory Grant NAGW-2422, NSF Grants 90-18526, ASC 93-18185 and AST 91-17388, and the Presidential Faculty Fellows program.

| Model | $M$ | $r_0$ | $N$ | Orbit type | $r_{tidal}$ | Impulse |
|-------|-----|-------|-----|------------|-------------|---------|
| A | $10^7$ | 0.6 | $10^5$ | 1 | 0.36 | 3.67 |
| B | $10^8$ | 0.6 | $10^5$ | 1 | 0.77 | 0.37 |
| C | $10^8$ | 0.6 | $10^5$ | 2 | 0.82 | 0.26 |
| D | $10^8$ | 0.6 | $5 \times 10^6$ | 3 | 0.77 | 0.37 |

Table 1: Parameters of simulations. Length scales are in kpc and mass is in $M_\odot$. $N$ is the number of particles used in the simulation and the impulse is expressed in units of the initial binding energy of the dwarf galaxy.

| Orbit type | Pericenter | Apocenter | Period |
|------------|------------|-----------|--------|
| 1 | 13.4 | 81.5 | 1.08 |
| 2 | 15.5 | 81.5 | 1.18 |
| 3 | 14.4 | 185.4 | 2.48 |

Table 2: Parameters of orbits. Length is in kpc, time is in Gyrs.

# REFERENCES

Aguilar, L., Hut, P. & Ostriker, J.P. 1988, ApJ, 335, 720

Bahcall, J.N., Flynn, C., Gould, A. & Kirhakos, S. 1994, preprint

Bergbusch, P.A. & VandenBerg, D.A. 1992, ApJS, 81,163

Binney, J. & Tremaine, S. 1987, Galactic Dynamics (Princeton University Press, Princeton)

Bolte, M. 1994, PASP, in preparation

Buonanno, R., Corsi, C.E. & Fusi Pecci, F. 1985, A&A, 145,97

Gerhard, O.E. 1993, MNRAS, 265, 213

Green, P.J., Margon, B., Anderson, S.F. & Cook, K.H. 1994, ApJ, 434, 319

Hernquist, L. 1990a, ApJ, 356,359

Hernquist, L. 1990b, in Warped Disks and Inclined Rings Around Galaxies p.96, ed. Casertano, S., Sackett, P. & Briggs, F. (Cambridge University Press, Cambridge)

Hernquist, L. & Barnes, J. E. 1990, ApJ, 349, 562

Hernquist, L. & Ostriker, J. P. 1992, ApJ, 386, 375

Hernquist, L. & Quinn, P.J. 1993, in Galaxy Evolution: the Milky Way Perspective, p.187, ed. S.R. Majewski (Astronomical Society of the Pacific, San Francisco)

Hernquist, L., Sigurdsson, S. & Bryan, G. 1994, ApJ, submitted

Heyl, J., Hernquist, L. & Spergel, D.N. 1994, ApJ, submitted

Hodge, P. 1988 PASP, 100, 568

Ibatha, R.A., Gilmore, G. & Irwin, M.J. 1994, Nature, 370, 194

Johnston, K.V. & Hernquist, L. 1994, in preparation

King, I.R. 1962, AJ, 67, 471

Larson, R.B. 1990, PASP, 102, 709

Lin, D.N.C. & Richer, H.B. 1992, ApJ, 388, L57

Lynden-Bell, D. 1976, MNRAS, 174, 695

Lynden-Bell, D. 1982, Observatory, 102, 202

Majewski, S.R. 1993a, ARA&A, 31, 575

Majewski, S.R. 1993b, in Galaxy Evolution: the Milky Way Perspective, p.5, ed. S.R. Majewski (Astronomical Society of the Pacific, San Francisco)

Majewski, S.R. 1994, ApJ, 431, L17

Majewski, S.R., Munn, J.A. & Hawley, S.L. 1994, ApJ, 427, L37

Mateo, M., Udalski, A., Szymański, M., Kałuzny, J., Kubiak, M. & Krzemiński, W. 1994, preprint

Mateo, M., Olszewski, E.W., Pryor, C., Welch, D.L. & Fischer, P. 1993, AJ, 105, 510

McGlynn, T.A. 1990, ApJ, 348, 515

Mighell, K. 1990, A&AS, 82, 207

Miller, R.H. 1986, A&A, 167, 41

Miyamoto, M. & Nagai, R. 1975, PASJ, 27, 533

Oh, K.S., Lin, D.N.C. and Aarseth, S.J. 1994, ApJ, in press

Preston, G.W., Beers, T.C. & Shectman, S.A. 1994, AJ, 108, 538

Quinn, P.J. and Goodman, J. 1986, ApJ, 309, 472

Quinn, P.J., Hernquist, L. and Fullagar, D.P. 1993, ApJ, 403, 74

Reid, N. & Majewski, S.R. 1993, ApJ, 409, 635

Richer, H.B. and Westerlund, B.E. 1983 ApJ, 264, 114

Searle, L. & Zinn, R. 1978, ApJ, 225, 357

Tóth, G. & Ostriker, J.P. 1992, ApJ, 389, 5

van der Marel, R.P. & Franx, M. 1993, ApJ, 407, 525.

Walker, I., Mihos, J.C. & Hernquist, L. 1994, in preparation

Weinberg, M. 1993, ApJ, 410, 543

# Figure Captions

**Figure 1** Orbits in the X-Z plane. The disk of the Milky Way lies in the X-Y plane.

**Figure 2** Fractional mass remaining bound as a function of time in orbital periods (T) for each simulation. The "Fig" labels show the times at which the state of a model is presented in the following Figs 3 – 7.

**Figures 3 - 7** Summary of analysis at times indicated in Figure 2.
**(a)** Surface density contours. The bold contour represents $1.0 M_\odot/\text{pc}^2$. Consecutive contours are a factor of $\sqrt{10}$ apart.
**(b)** Line of sight average velocity (open symbols) and velocity dispersion (closed symbols) along the projected major axis.
**(c)** Normalized Gauss-Hermite moments (see text).
**(d)** Line of sight velocity distribution of particles in the bin at $-1°$ (used in (b) and (c); f is the fraction of particles per unit $\sigma$). The solid line is a Gaussian of the same velocity dispersion. The dotted line includes the higher Gauss-Hermite moments. The crosses indicate the points from simple binning of the data.
**(e)** As (f), but for the $-9°$ bin.
**(f)** Surface density profile (solid line). For comparison, a King model of the given concentration (c= $log(r_{tidal}/r_{core})$) is shown by the dotted line.
**(g)** Point along orbit at which the analysis was done.

**Figure 8** Galactocentric radial velocity vs. radius for random sample of 2500 particles. The solid line is the locus of the orbit in this plane.

**Figure 9** Analysis of tidal streamers. N is the number of stars per (degree)$^2$ per 1 km/s bin at 40° (panels (a) and (b)) and 75° (panels (c) and (d)) from the dwarf. The dotted line shows the background towards the North Galactic Pole. The solid line is from the simulation. The different panels are for the stated magnitude bins.

**Figure 10** As figure 9, but for the dwarf at 0.5 Gyrs (panels (a) and (b)) and 1.0 Gyrs (panels (c) and (d)) after disruption.